\documentclass[12pt, draftclsnofoot, onecolumn]{IEEEtran}
\usepackage{amsfonts}
\usepackage{slashbox}
\usepackage{amsmath}
\usepackage{multirow}
\usepackage{graphicx}
\usepackage{amsfonts}
\usepackage{amsmath,epsfig}
\usepackage{mathbbold}
\usepackage{stfloats}
\usepackage{amssymb}
\usepackage{url}
\usepackage{bm}
\usepackage{cite}
\usepackage{amsmath}
\usepackage{amssymb}
\usepackage{latexsym}
\usepackage{multirow}
\usepackage{epsfig}
\usepackage{graphics}
\usepackage{mathrsfs}
\usepackage{algorithmic}
\usepackage{algorithm}
\usepackage{subfigure}
\usepackage{slashbox}
\usepackage[all]{xy}

\usepackage{pifont}
\usepackage{bbding}
\usepackage{stmaryrd}
\usepackage{amssymb}
\usepackage{amsfonts}
\usepackage{epic}
\usepackage{stfloats}
\usepackage{latexsym}
\usepackage{epstopdf}
\usepackage{epic}
\usepackage{bm}
\usepackage{xcolor}
\usepackage{mathrsfs}
\usepackage{pifont}
\usepackage{bbding}
\usepackage{amsmath,epsfig}
\usepackage{mathbbold}
\usepackage{stmaryrd}
\usepackage{amssymb}
\usepackage{amsfonts}
\usepackage{epic}
\usepackage{graphicx}
\usepackage{subfigure}

\usepackage{enumerate}

\usepackage{stfloats}
\usepackage{latexsym}
\usepackage{epstopdf}
\usepackage{epic}
\usepackage{multirow}
\usepackage{stfloats}
\usepackage{bm}

\usepackage{booktabs}
\usepackage{color}
\usepackage{cases}
\usepackage{float}

\begin{document}
\title{Fundamental Tradeoffs in Communication and Trajectory Design for UAV-Enabled Wireless Network}
\author{\IEEEauthorblockN{Qingqing Wu, \emph{Member, IEEE}, Liang Liu, \emph{Member, IEEE}, and Rui Zhang, \emph{Fellow, IEEE}
\thanks{ The authors are with the Department of Electrical and Computer Engineering, National University of Singapore, email:\{elewuqq, eleliu, elezhang\}@nus.edu.sg.}} }

\maketitle
\begin{abstract}
The use of unmanned aerial vehicles (UAVs) as aerial communication platforms is of high practical value for future wireless systems such as 5G, especially for swift and on-demand deployment in temporary events and emergency situations. Compared to  traditional terrestrial base stations (BSs) in cellular network, UAV-mounted aerial BSs possess stronger line-of-sight (LoS) links with the ground users due to their high altitude as well as high and flexible mobility in three-dimensional (3D) space, which can be exploited to enhance the communication performance.  On the other hand, unlike terrestrial BSs that have reliable power supply, aerial BSs in practice have limited on-board energy, but require
significant propulsion energy to stay airborne and support high mobility.
Motivated by the above new considerations, this article aims to revisit some fundamental tradeoffs in UAV-enabled communication and trajectory design. Specifically, it is shown that communication throughput, delay, and (propulsion) energy consumption can be traded off among each other by adopting different UAV trajectory designs, which sheds new light on their traditional tradeoffs in terrestrial communication. Promising directions for future research are also discussed.
\end{abstract}

\section{Introduction}\label{sec:Introduction}
\begin{figure}[!t]
\centering
\scalebox{0.39}{\includegraphics*{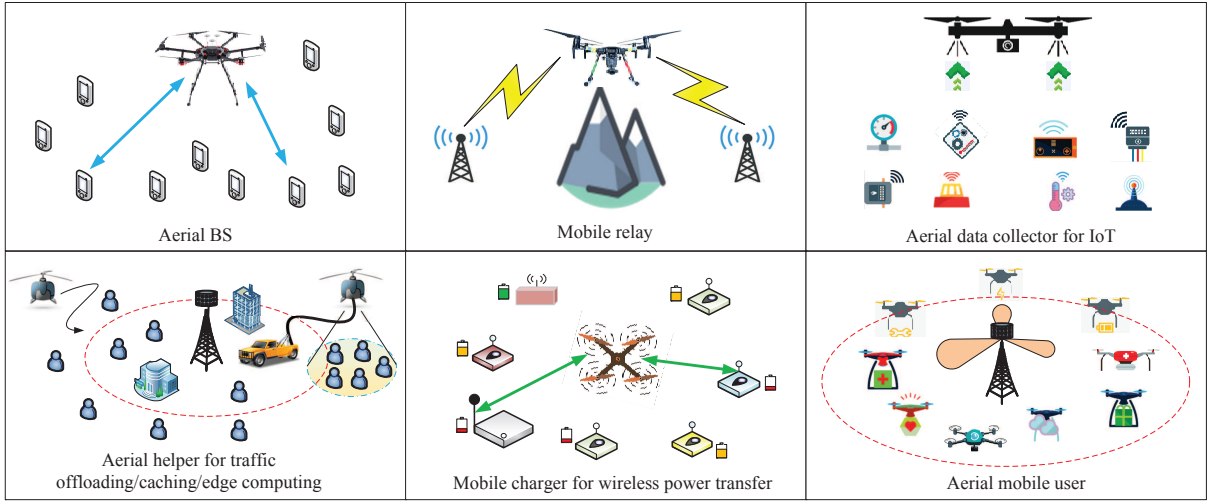}}
\caption{Typical UAV communication applications in 5G network. } \label{UAV:applications}
\end{figure}

Due to their prominent features of high mobility and flexible deployment, unmanned aerial vehicles (UAVs), also known as drones, have found many promising usages in the forthcoming fifth-generation (5G) and future beyond-5G wireless networks, as shown in Fig. \ref{UAV:applications}. Particularly, UAVs can be used cost-effectively as on-demand aerial platforms to provide or enhance the communication services for terrestrial mobiles/devices in a multitude of applications, including aerial base stations (BSs)/relays in situations without the terrestrial cellular coverage \cite{Guvenc16,Zhang16,Gesbert17}, aerial helpers for providing new services such as data backhaul/offloading, cached-content multicasting and edge computing for terrestrial BSs/users, and mobile hubs for energy-efficient data collection \cite{mozaffari2017mobile} and wireless power transfer \cite{xu2017uav2} for low-power Internet-of-things (IoT) devices such as sensors and tags. On the other hand, UAVs in many civilian applications such as cargo delivery and aerial video surveillance may become  new aerial users in the cellular network, which need to have high-performance two-way communications with the ground BSs to receive control signals and upload payload data in real time \cite{zhang2017cellular}.


Despite the above promising UAV applications, their future success critically depends on the development of new and effective  UAV-to-ground communication technologies. Compared to the conventional terrestrial communications, UAV-to-ground communications enjoy the following two main advantages that can be exploited for throughput enhancement, namely line-of-sight (LoS) dominated UAV-to-ground channels and UAV's controllable high mobility in three-dimensional (3D) space. On one hand, thanks to the high altitude of UAVs, the probability of LoS channels between UAVs and the ground users/BSs is in general pretty high, and thus UAV-to-ground communications are significantly less affected by channel impairments such as shadowing and fading compared to terrestrial communications. On the other hand, thanks to the high mobility, swift 3D deployment or even dynamic movement of UAVs becomes feasible so that they can adjust their locations/trajectories based on the locations and/or movement of the ground BSs/users to maintain favorable LoS channels. It is worth noting that the LoS channels enable UAVs to have their signal coverage over a much larger number of ground users or BSs as compared to the BSs/users  in terrestrial communications. Consequently, to achieve optimal communication and trajectory design, each UAV should not only maintain strong channels to its served users or connected BSs via flying in proximity of them, but also control its interferences to other UAVs as well as ground users/BSs so as to maximize the network throughput.

\begin{figure}[!t]
\centering
\subfigure[Throughput-delay tradeoff.]{\includegraphics[width=2.1in, height=1.9in]{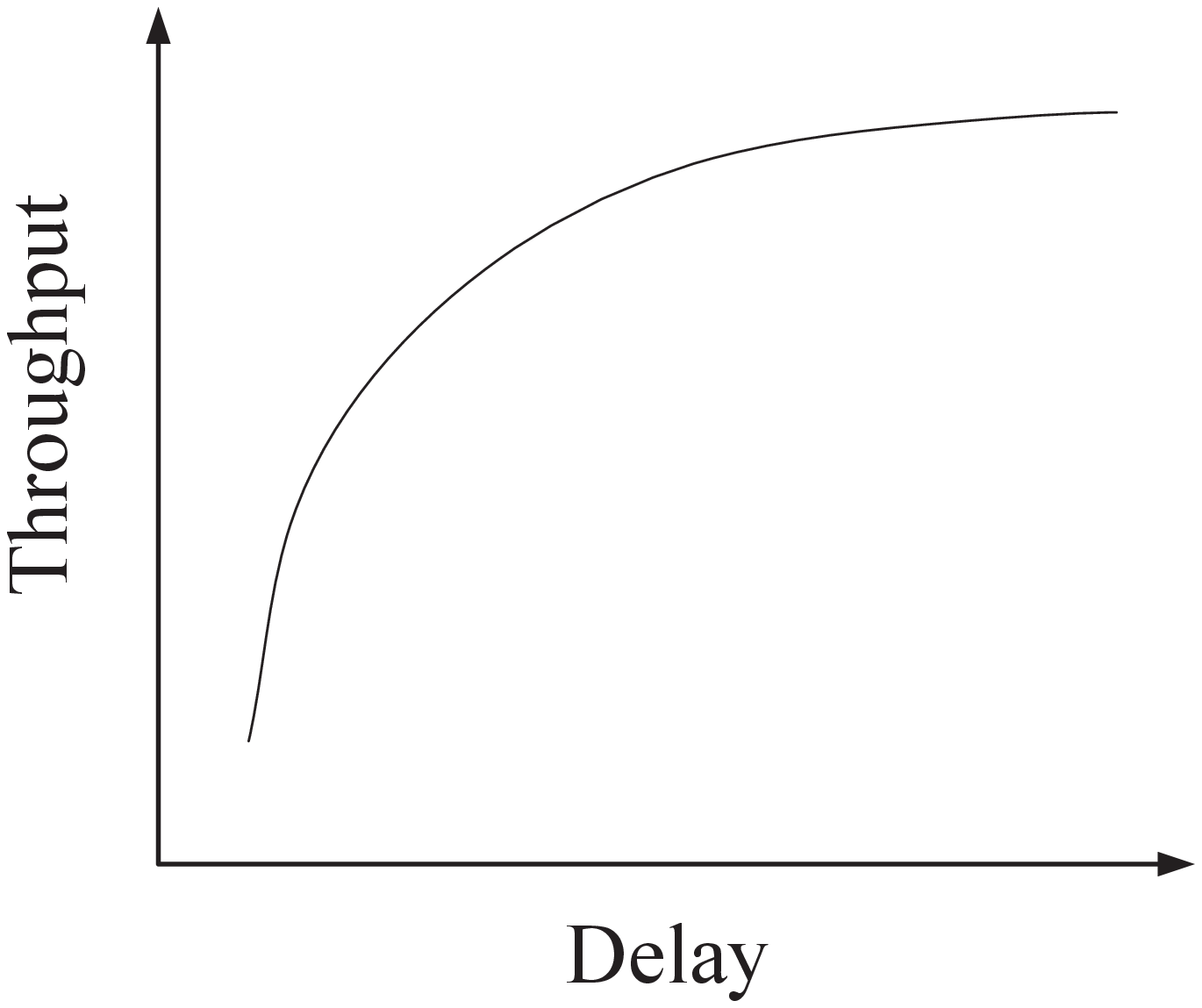}}~
\subfigure[Throughput-energy tradeoff.]{\includegraphics[width=2.1in, height=1.9in]{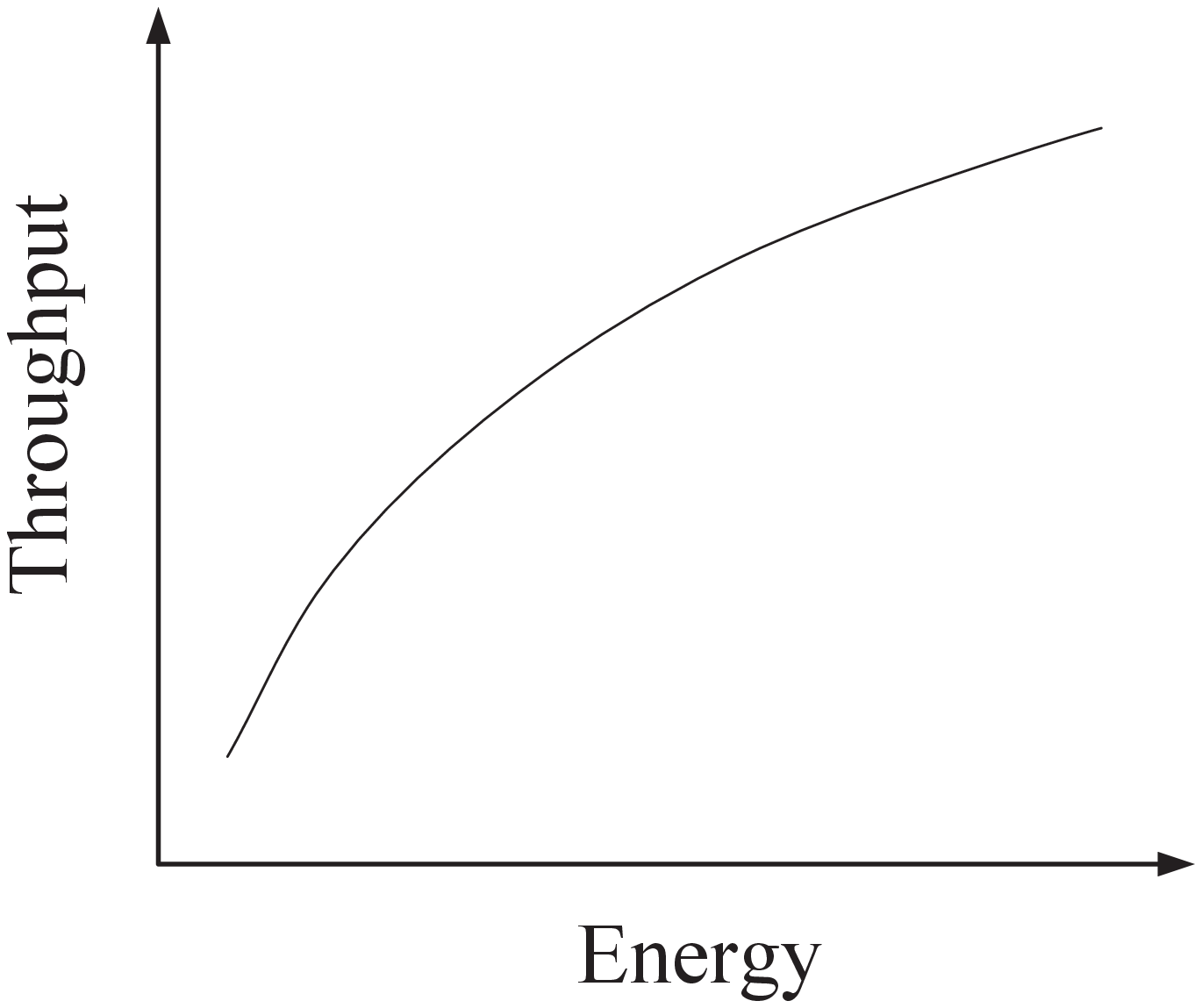}}
\subfigure[Delay-energy tradeoff.]{\includegraphics[width=2.1in, height=1.9in]{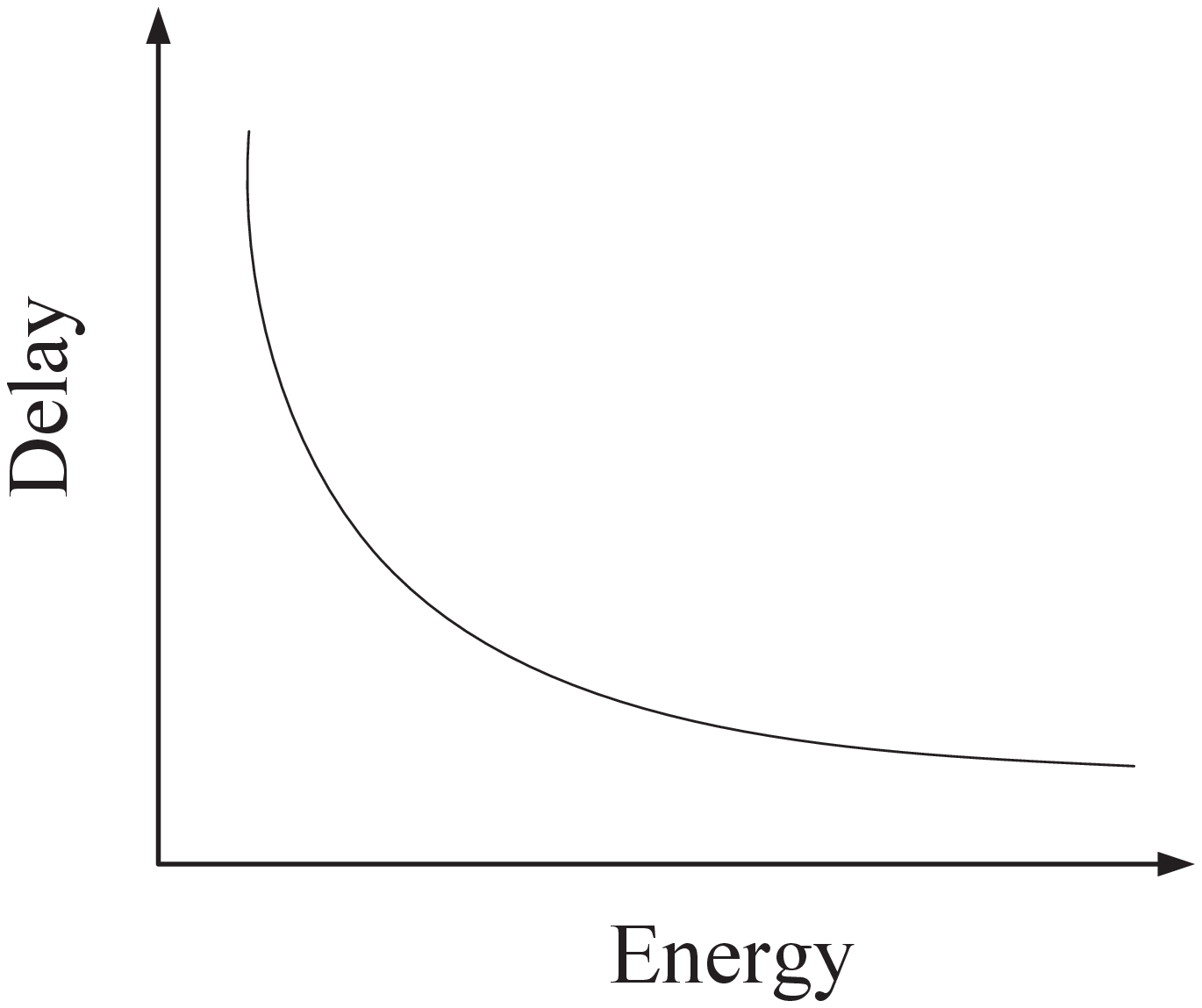}}
\caption{Three fundamental tradeoffs in UAV-enabled wireless communication.} \label{TDE:tradeoff:curve}\vspace{-0.1cm}
\end{figure}

Besides throughput, two important factors also need to be considered in UAV communication and trajectory deign, namely delay and energy. First, to maximize throughput, each UAV should communicate with a ground user/BS when its flies sufficiently close to them so as to reduce their distance and hence improve the link capacity. However, this inevitably incurs more delay in communication due to the UAV movement. Thus, there is an interesting throughput-delay tradeoff in UAV-to-ground communications, as shown in Fig. \ref{TDE:tradeoff:curve} (a). Second, there also exists a new tradeoff between throughput and energy in UAV-enabled communication as shown in Fig. \ref{TDE:tradeoff:curve} (b), since the UAV generally needs to consume more propulsion energy to move closer to the ground users/BSs in order to gain higher throughput. Last, the above two tradeoffs naturally imply a delay-energy tradeoff as shown in Fig. \ref{TDE:tradeoff:curve} (c), as delay in UAV-to-ground communication can be reduced if more propulsion energy is consumed by the UAV to move faster to the ground users/BSs it is designated to communicate with.

Motivated by the above new and interesting tradeoffs between throughput, delay and (propulsion) energy consumption in UAV communication and trajectory design, this article aims to provide an overview on the state-of-the-art results in this new area. In particular, we will focus on the use of UAVs as communication platforms (e.g., aerial BSs/relays) to serve the terrestrial users, although such fundamental tradeoffs also exist similarly in the other paradigm with UAVs as new aerial users to be  served by the ground BSs in the cellular network. Next,  we discuss the main differences between these tradeoffs in UAV-enabled communication and their counterparts in traditional terrestrial communication.

\section{Fundamental Tradeoffs in UAV-Enabled Communication}

It is well-known that there exist fundamental tradeoffs between the throughput, delay and energy in wireless communication \cite{tse2005fundamentals}. In this section, we first review the classic results on these tradeoffs in terrestrial communication, and then explain their main differences in UAV-to-ground communication arising from its unique characteristics as discussed in the preceding section, such as the LoS channel, the UAV's trajectory design and its high propulsion energy consumption.

\subsection{Throughput-Delay Tradeoff}\label{sec:Throughput-Delay Tradeoff}
The throughput-delay tradeoff has been extensively studied for the terrestrial wireless communication. For a basic point-to-point wireless communication link, the maximum achievable rate over fading channels, defined as the ergodic capacity, is achieved by coding over a sufficiently large number of channel coherence intervals to fully exploit the ergodicity of fading channels \cite{tse2005fundamentals}. However, this comes at the cost of long transmission delay that may not be tolerable for applications with stringent latency requirement. On the other hand, channel coding can be performed over each coherence interval to reduce the delay, resulting in the so-called delay-limited capacity \cite{tse2005fundamentals}. However, delay-limited capacity is in general smaller than the ergodic capacity, and outage is usually inevitable in deep fading. For the general multi-user communication, the multi-user diversity gain can be attained to improve the network throughput by scheduling the user with the best channel among all users to communicate in each coherence interval, whereas this inevitably leads to more significant delay for each user as the number of users increases \cite{tse2005fundamentals}. The above results show that there is a general throughput-delay tradeoff for communication over fading channels.  Moreover, it is shown in \cite{Tse02} that there is another tradeoff between the total throughput of a mobile ad-hoc network  (MANET) and the average delay tolerable by the users in the network due to the random user movement, as each user can wait to communicate with another user until they become sufficiently close to each other.

By contrast, in UAV-enabled communication, channel fading is no longer a key factor contributing to the throughput-delay tradeoff thanks to the LoS-dominated channels. Instead, the mobility of UAVs plays the deciding role in such tradeoff as the UAV-to-ground LoS channels are solely determined by the distances between the UAV and ground users, which critically depend on the UAV location. However, in sharp contrast to the random user movement in a MANET where the delay is random and difficult to predict for the users  \cite{Tse02}, the delay in UAV-enabled communication can be properly controlled via joint UAV trajectory and communication scheduling design. Moreover, another key difference lies in the time scale of the delay between the terrestrial communication and UAV-to-ground communication: in the former case, the delay is measured in terms of channel coherence time, e.g., milliseconds, while in the latter case, the delay is proportional to the UAV flying time (distance divided by speed), e.g., seconds. As a result, in order to exploit the throughput-delay tradeoff via trajectory design in UAV-enabled communication, the application needs to be more delay tolerant as compared to that in terrestrial communication.

\subsection{Throughput-Energy Tradeoff}\label{sec:Throughput-Energy Tradeoff}
The throughput-energy tradeoff in the traditional wireless  communication is fundamentally rooted in the Shannon capacity formula which explicitly suggests that the achievable rate increases monotonically with the transmit power \cite{tse2005fundamentals}. One useful performance metric stemming from this tradeoff is energy efficiency, which indicates how many information bits can be transmitted using a Joule energy. If only the transmit energy is considered, it is well-known that the energy efficiency monotonically increases with decreasing  the transmit rate \cite{tse2005fundamentals}; while if the circuit power at the transmitter is considered as well, it is shown in \cite{miao2010energy} that the energy efficiency first increases and then decreases with the transmit rate.

In UAV-enabled communication, the propulsion energy (usually in the order of kilowatt (kW)) required to maintain the UAVs airborne and support their high mobility is generally several orders of magnitude higher than the transmit and circuit energy (usually in the order of watt (W)  or even smaller). As a result, the effect of propulsion energy on the UAV trajectory is the dominant factor determining  the throughput-energy tradeoff in UAV-enabled communication. For example, to enhance the throughput, each UAV needs to fly over a longer distance with a faster speed so that it can reach each of its served ground users as close as possible and stay near them as long as possible, given a finite flight duration. Moreover, each UAV also needs to adjust its altitude and/or make sharp turns to avoid blockages in the directions of its served ground users. All these can lead to more significant propulsion energy consumption for throughput enhancement.

\subsection{Delay-Energy Tradeoff}\label{sec:Delay-Energy Tradeoff}
As discussed in the above two subsections, the throughput-delay and throughput-energy tradeoffs in UAV-enabled communication exhibit interesting new aspects compared to their traditional counterparts in terrestrial communication. As a result, their corresponding delay-energy tradeoffs are also drastically  different due to the new UAV trajectory design and its high  propulsion energy consumption. For example, to reduce delay, each UAV should fly among its served ground users with its maximum speed, but remain at its minimum speed (e.g., hovering)  when serving them in its proximity, both resulting in  more propulsion energy consumption.

In the rest of this article, we will focus on examining the throughput-delay and throughput-energy tradeoffs, in the next two sections, respectively. We will provide concrete examples to illustrate them more clearly, provide overviews on their state-of-the-art results, and also point out promising directions for future research.

\section{Throughput-delay Tradeoff}\label{sec:Throughput-delay Tradeoff}
In this section, we investigate the joint UAV trajectory and communication design to characterize  the throughput-delay tradeoff. Specifically, we first consider a simple setup  with one UAV serving two GUs to draw useful  insights. Then, we extend our study to the general case with multiple UAVs serving multiple  users, followed by further discussions on related/future work.

\subsection{Single-UAV Enabled Wireless Network}\label{Sect:singleUAV}
\begin{figure}[!t]
\centering
~~~~~\,\subfigure[A UAV-enabled two-user wireless system.]{\includegraphics[width=2.8in, height=2.1in]{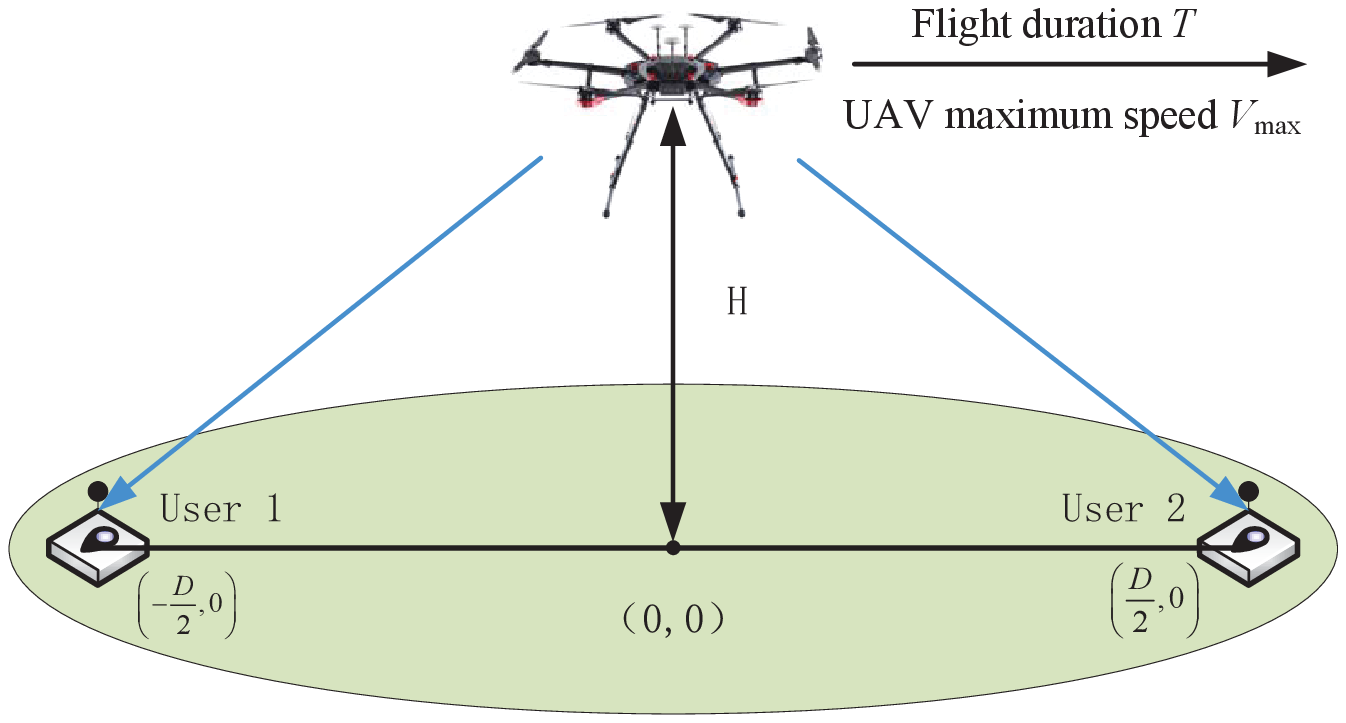} \label{singleUAV:model}}~
\,\,~~\subfigure[UAV's horizontal trajectory for different $T$.]{\includegraphics[width=3.2in, height=2.5in]{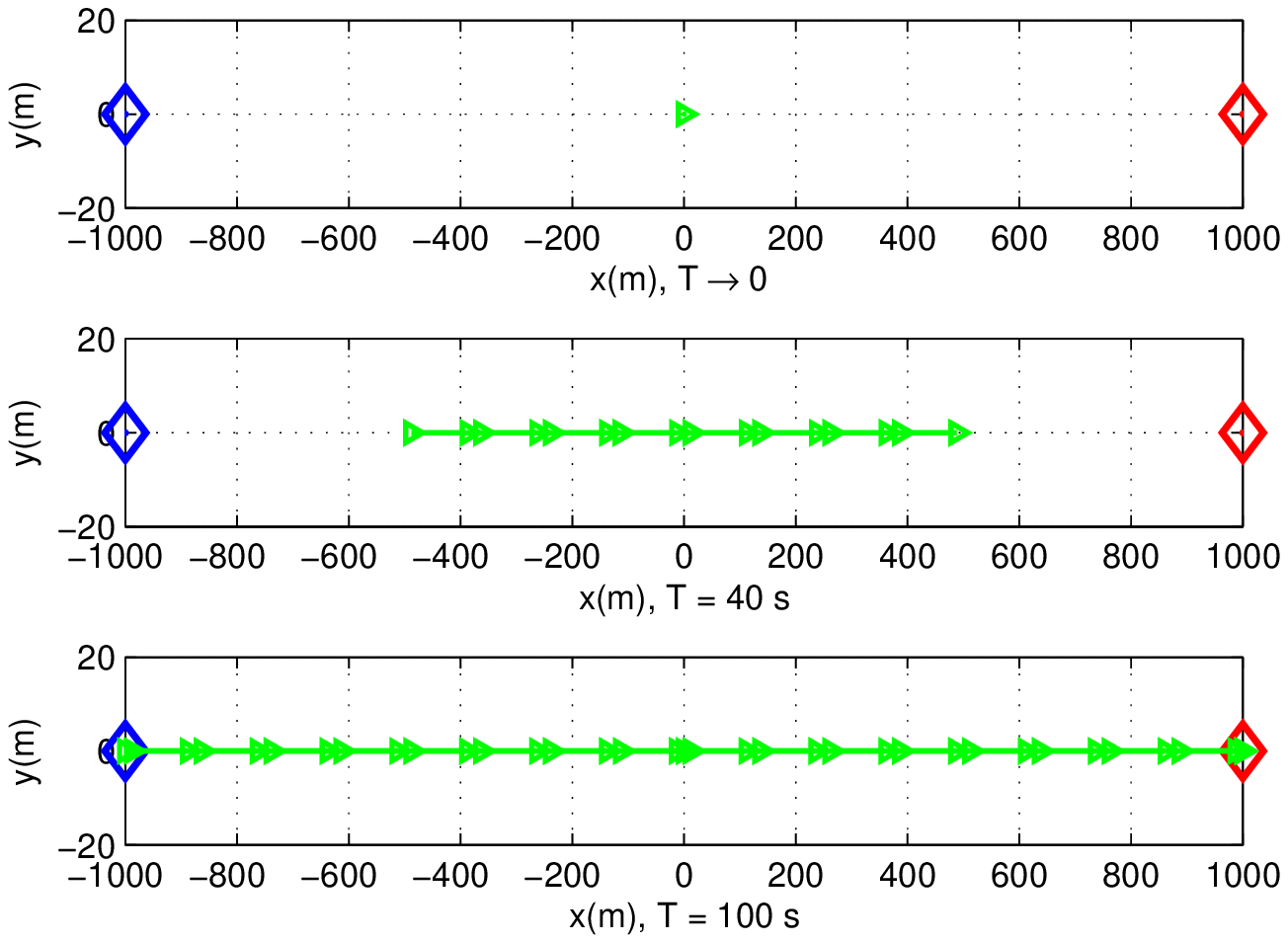}\label{TD:trj}}
\subfigure[Periodic TDMA  of GUs.]{\includegraphics[width=3.2in, height=2.5in]{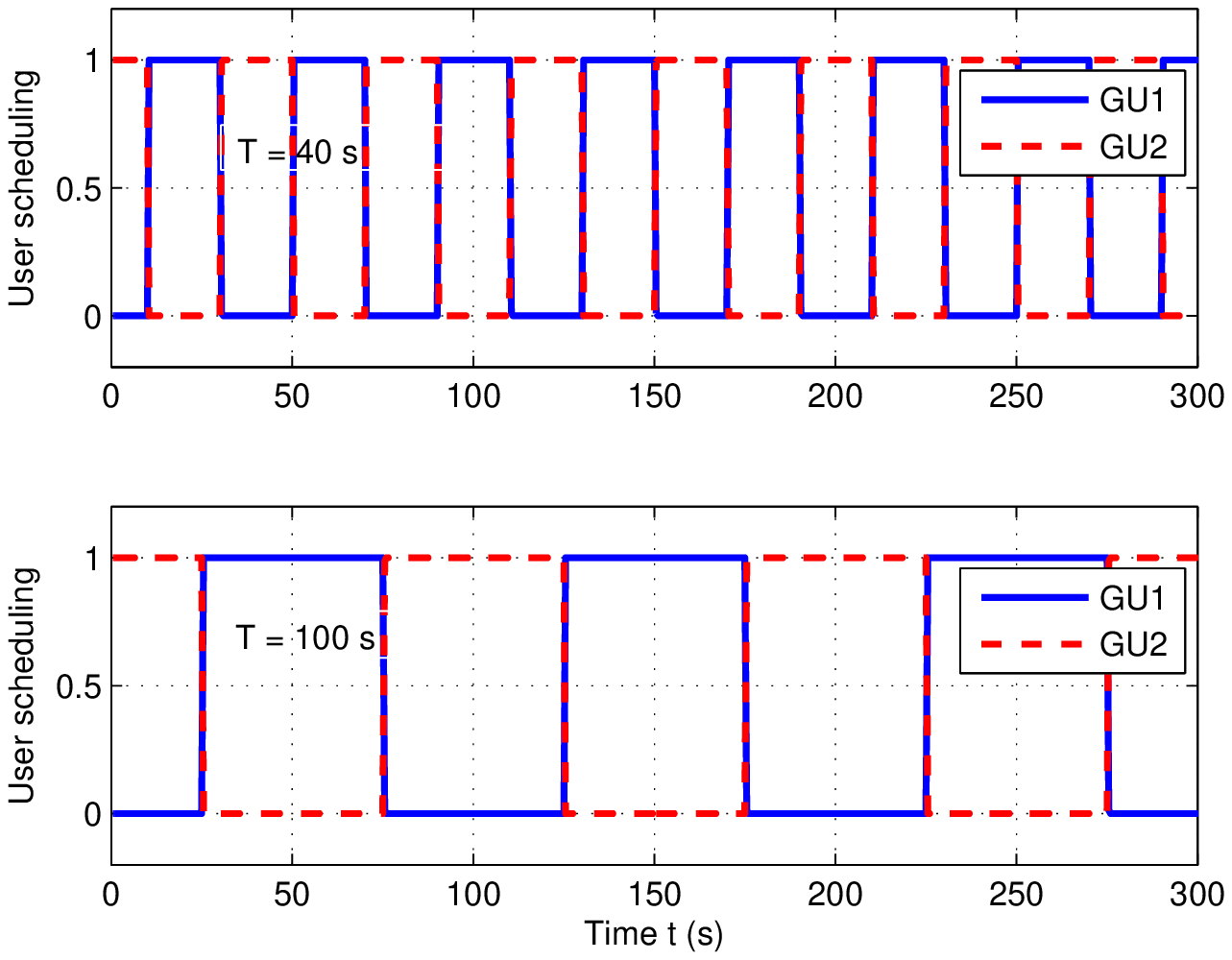}\label{TD:delay}}
\subfigure[Common throughput versus UAV flight period, $T$.]{\includegraphics[width=3.2in, height=2.5in]{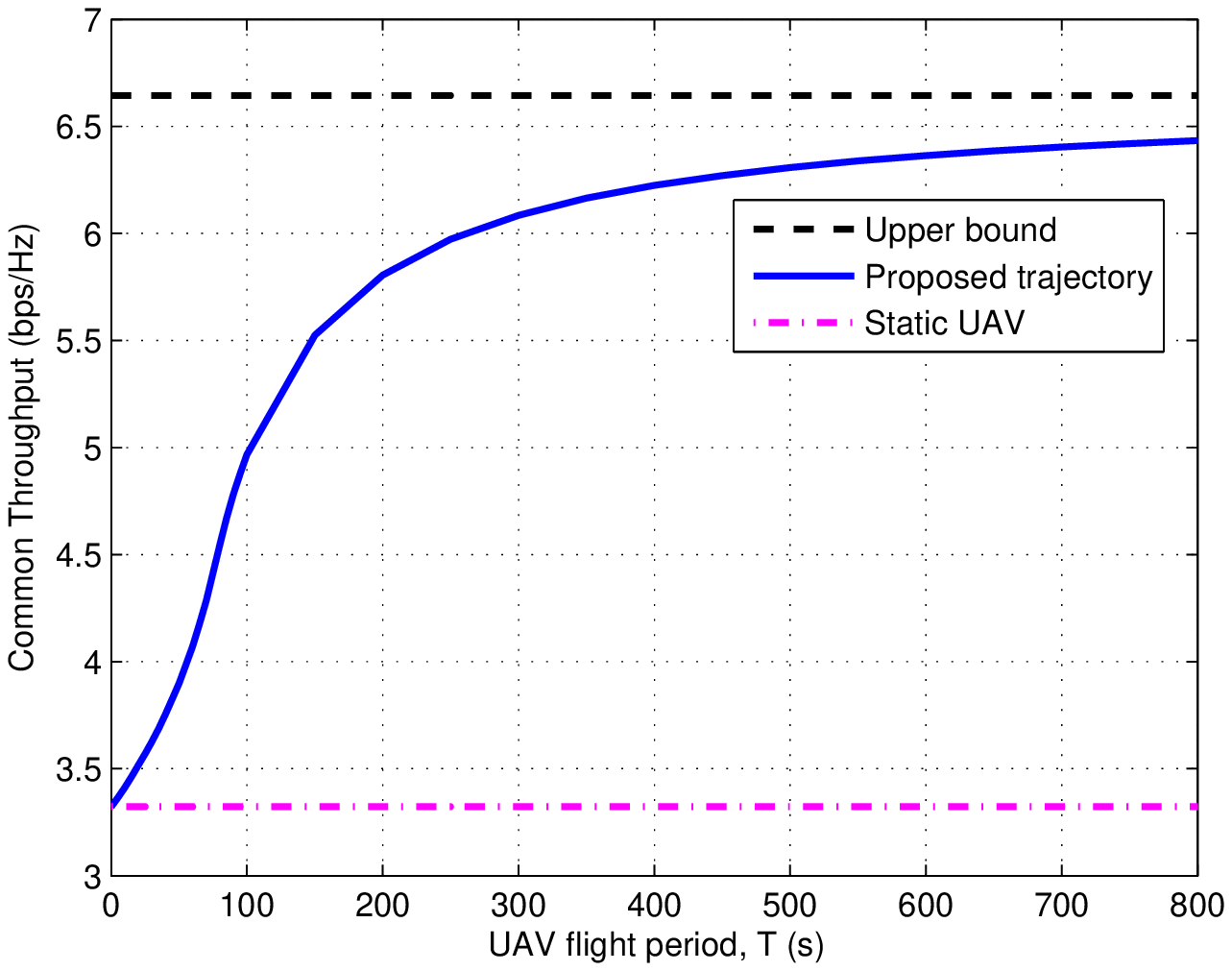}\label{TD:curve}}  
\caption{Throughput-delay tradeoff  for a single UAV-enabled network with two GUs.  The GUs' nominal locations are marked by `$\diamondsuit$'s and the UAV trajectories are marked by `$\vartriangleright$'s. The receiver noise power and the channel power gain at the reference distance 1 m are set as $-110$ dBm and $50$ dB, respectively, and the maximum transmit power is set as 0.1 W. Other required parameters are set as follows:  $V_{\max}=50$ m/s, $H=100$ m, and $D=2000$ m. } \label{Throughput:delay:tradeoff:singleUAV}\vspace{-0.5cm}
\end{figure}

As shown in Fig. \ref{singleUAV:model}, we consider a UAV-enabled wireless system where one UAV is employed to serve two GUs in a finite period of $T$ seconds (s).
 The UAV is assumed to fly at a constant altitude of $H$ in meter (m) with the maximum allowable speed denoted by $V_{\max}$ in meter/second (m/s). The air-to-ground channels from the UAV to GUs are assumed to be dominated by the LoS links. As such,  it is preferable to let the UAV fly as low as possible in order to reduce the signal path loss with the GUs. However, the minimum value of $H$ is practically limited for terrain or building avoidance.  The two GUs are assumed to be quasi-stationary with a distance of $D$ m between their nominal locations, as shown in Fig. 3(a),  where we assume that their maximum movement distances from their respective nominal locations within  the given period $T$ are negligible compared to  $D$ and the UAV altitude $H$; thus, their effects to the corresponding LoS channel gains are ignored.  We consider that the UAV communicates with GUs via  time-devision multiple access (TDMA), i.e., only one GU is scheduled for communication at any time instant, while other multiple access schemes will be discussed in Section \ref{MA:discussion}.  To serve GUs continuously in a periodic manner, we assume that the UAV needs to return to its initial location by the end of each flight period $T$ while the initial location can be optimized for maximizing the throughput.  As such, the delay of each GU in the worst case is $T$ s.
To ensure fairness among GUs, we aim to maximize the common (minimum) throughput between among the GUs via jointly optimizing the UAV trajectory and communication.

In Fig. \ref{TD:trj}, we show the optimal UAV's horizontal trajectories projected onto the horizontal ground plane under different flight periods, $T$. It can be observed that as $T$ increases, the UAV tends to fly closer to  the two GUs while when $T$ is sufficiently large (e.g., $T = 100$ s), the UAV  flies between the two GUs with its maximum speed to save more time for hovering right above the two GUs to maintain the best channel for communication.
Furthermore, at any time instant, to maximize the throughput, the GU that is closer to the UAV (thus with a better channel) should be scheduled for communication, while the other GU has to wait until the UAV flies closer to it again. As such,  each GU will experience a waiting time of  $T/2$ s for communicating with the UAV periodically. This is illustrated in Fig. \ref{TD:delay} where the user scheduling is plotted over time.  It is observed that a larger $T$ leads to a longer waiting time for each GU.
Finally, we show in Fig. \ref{TD:curve} the achievable common throughput in bits per second per Hertz (bps/Hz) versus  $T$. Note that the throughput upper bound  is obtained by ignoring the time spent on travelling between the two GUs, which holds when $T$ goes to infinity.  In addition, the throughout of a static UAV is obtained by fixing the UAV at the middle location between the two GUs for all time.  One can observe that compared to the case of a static UAV,  the common throughput can be significantly improved as  $T$ increases with a mobile UAV. However,  such a throughput gain is at the cost of scarifying the user delay, which thus reveals  a fundamental throughput-delay tradeoff in UAV-enabled wireless network.

\subsection{Multi-UAV Enabled Wireless Network}\label{sec:Multi-UAV Enabled Wireless Network}
 The use of multiple UAVs for cooperatively serving the GUs is an effective solution to improve the throughput-delay tradeoff over the single-UAV enabled network, by dividing the GUs into smaller groups, each served by one of the UAVs.
\begin{figure}[!t]
\centering
~~~~~\,\subfigure[A multi-UAV enabled wireless system with IUIC.]{\includegraphics[width=2.8in, height=2.1in]{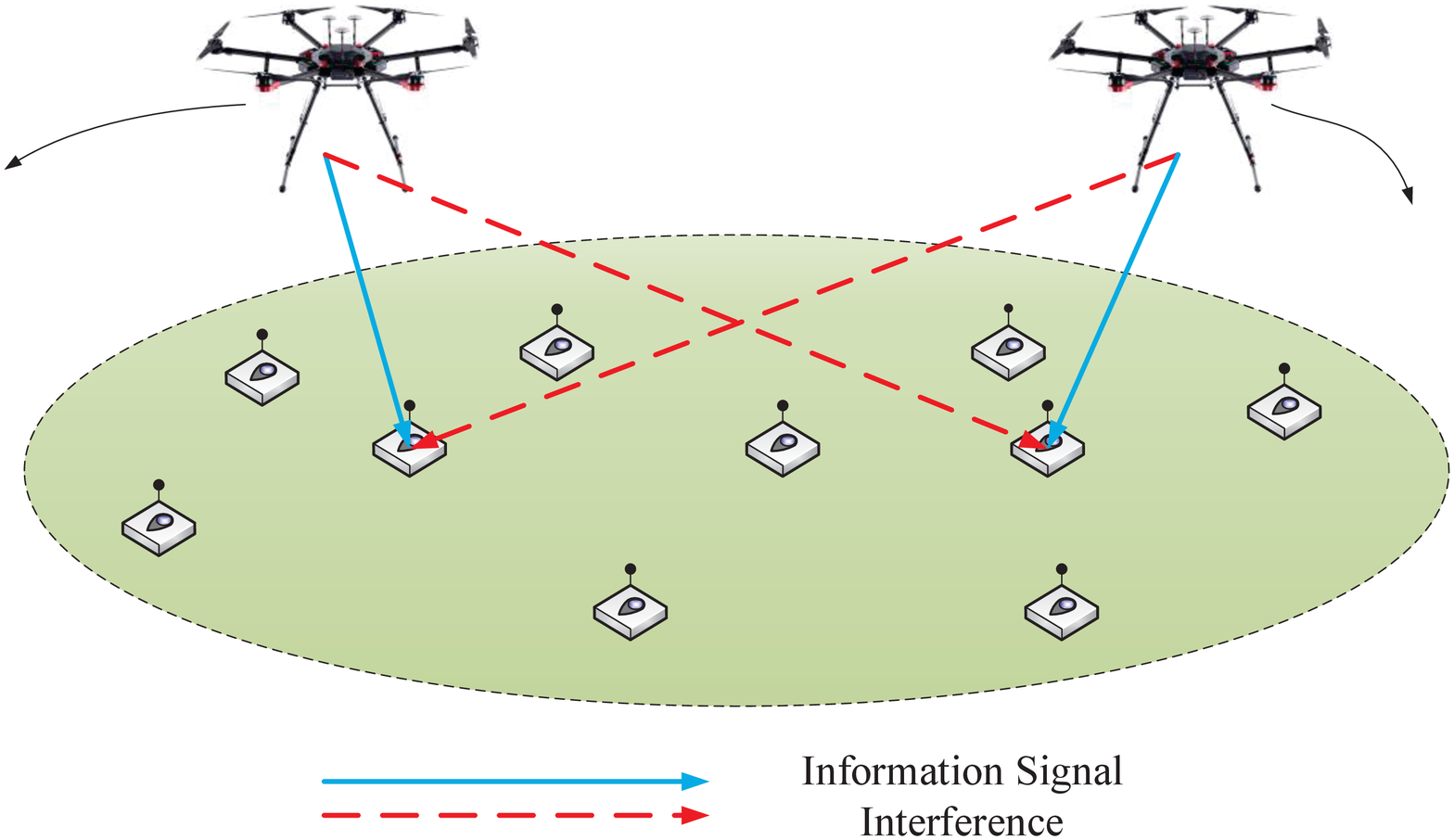} \label{multiUAV:model}}~
\,\,~~\subfigure[UAV horizontal trajectories without power control.]{\includegraphics[width=3.2in, height=2.5in]{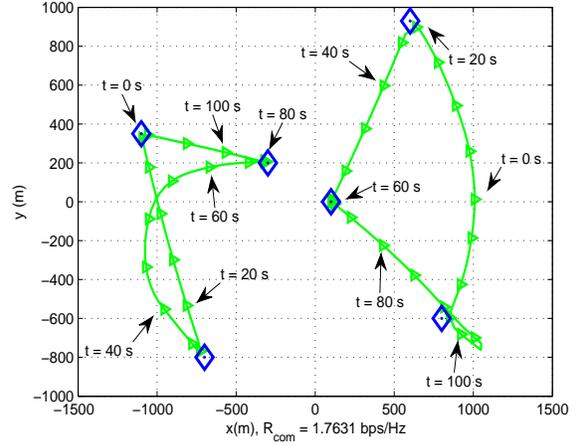}\label{multiUAV:trj:nopower}}
\subfigure[UAV horizontal  trajectories with power control.]{\includegraphics[width=3.2in, height=2.5in]{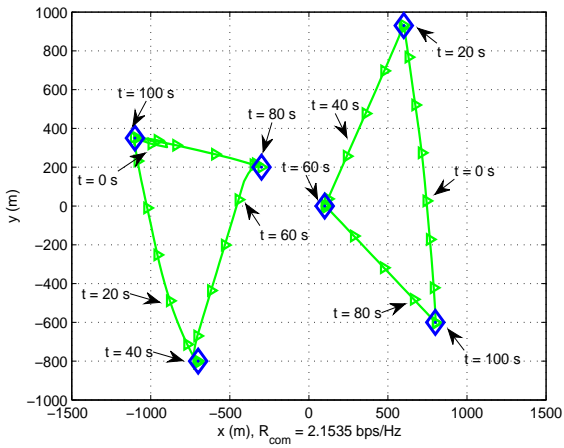}\label{multiUAV:trj:power}}
\subfigure[Common throughput versus UAV flight period, $T$.]{\includegraphics[width=3.2in, height=2.5in]{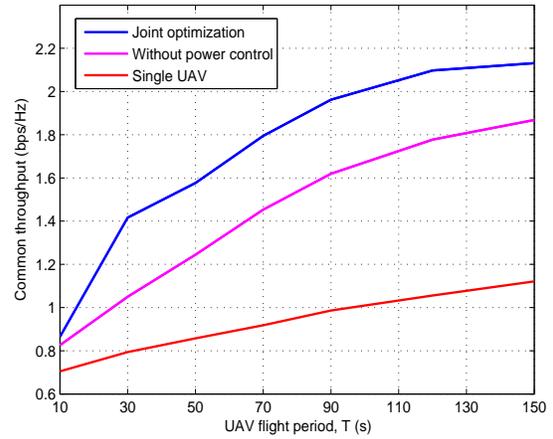}\label{multiUAV:TD:curve}}  
\caption{Throughput-delay tradeoff for a multi-UAV enabled wireless system with IUIC.  The GUs' nominal locations are marked by `$\diamondsuit$'s and the UAV trajectories are marked by `$\vartriangleright$'s.  The simulation parameters are set the same as those in Fig. \ref{Throughput:delay:tradeoff:singleUAV}. The user common throughput is denoted by $R_{\rm com}$ in bps/Hz.  } \label{multiUAV}\vspace{-0.5cm}
\end{figure}
 To demonstrate this, we consider a multi-UAV enabled system as shown in Fig. \ref{multiUAV:model} where $M=2$ UAVs are employed to serve a group of $K$ GUs in a finite period $T$. To achieve high spectral efficiency, we consider a spectrum sharing system where the UAVs share the same frequency band for communication and each of the UAVs serves its associated GUs via the periodic TDMA.
   As such, each GU inevitably suffers from severe interference from the other UAV due to the LoS channel, which requires the  inter-UAV interference coordination (IUIC) by jointly designing the UAV trajectories, transmit power and user associations. Similar to Section \ref{Sect:singleUAV}, we maximize the common throughput of all GUs with optimized IUIC. However, this problem is a non-convex optimization problem involving infinite variables due to the continuous UAV trajectory. To tackle this problem, we first apply time discretization to divide the UAV flight period into a finite number of equal-time slots, each with a nominal location of the UAV. Then,  we apply the block coordinate descent and  successive convex optimization techniques to obtain a suboptimal solution to the IUIC design  \cite{wu2018joint}. As an initial UAV trajectory is required for our algorithm, we adopt a circular UAV trajectory for initialization \cite{wu2018joint}.

  For the purpose of illustration, we consider a setup with $K=6$ GUs. Specifically, we show the optimized UAV trajectories without and with power control in Figs. \ref{multiUAV:trj:nopower} and   \ref{multiUAV:trj:power}, respectively, for  $T=120$ s.  In the former case, both UAVs transmit with their maximum power at all time. It is observed from Fig.   \ref{multiUAV:trj:nopower}  that  the optimized UAV trajectories  not only shorten the communication distances between the UAVs and their associated GUs (e.g., from $t = 0$ to $t=20$ s),  but also enlarge  separations of the two UAVs  to help alleviate the co-channel interference (e.g., from $t = 40$ s to $t=60$ s), even without the use of power control. However, at certain  pairs of UAV locations, this is realized at the cost of sacrificing  direct link gains, especially when the UAVs  fly on the way to serve two GUs (e.g., the middle two GUs that are nearby) that are close to each other.

   In contrast, for the case with power control, it is observed from  Fig.   \ref{multiUAV:trj:power} that the optimized UAV trajectories do not tend to sacrifice the direct link gains in return for maximum distance separation. This is because power control can help avoid strong interference even when the two UAVs have to be close to each other (e.g., when serving the middle two GUs).
As a result,  the common throughput is substantially improved as shown in Fig. \ref{multiUAV:TD:curve}.  It is also observed that the user throughput in the multi-UAV network is significantly improved over the  single-UAV network at the same user delay, thus verifying the improved throughput-delay tradeoff via effective multi-UAV cooperation with IUIC.


%

\subsection{Further Discussion and Future Work}\label{MA:discussion}

Besides orthogonal multiple access schemes such as TDMA considered above, non-orthogonal multiple access schemes based on superposition coding (SC) or dirty paper coding \cite{tse2005fundamentals} can be jointly designed with the UAV trajectory to further improve the throughput-delay tradeoff and achieves the capacity limits of UAV-enabled wireless networks  \cite{JR:wu2017_capacity}. For example, a  two-user broadcast channel (BC) is studied in \cite{JR:wu2017_capacity}, where it is shown that a simple and practical hover-fly-hover (HFH) trajectory with SC achieves the capacity region.
However, whether similar results hold for a UAV-enabled BC with more than two users or other multiuser channel models still remains an open problem that is worth investigating in future work.
Furthermore, in our above discussion, the user delay is roughly measured by the UAV flight period. However,  the delay requirements in 5G networks may vary dramatically in time scale from milliseconds (e.g., online gaming/video streaming) to several seconds (e.g., file sharing/data collection). Thus, how to model such  delay requirements and design the joint UAV trajectory and communication resource allocation to meet them is also an important problem for future research.

In practice, the UAV-to-ground LoS channel model is appropriate for rural or sub-urban areas or when the UAV altitude is sufficiently high. However,  for other cases such as in urban environment, other air-to-ground channel models such as probabilistic LoS model and Rician fading model, may be more suitable. It is worth noting that such non-LoS channel models may have significant effects on the optimal UAV trajectory design in UAV-enabled wireless networks. For example, decreasing the UAV's flying altitude under the probabilistic LoS model generally decreases the probability of having LoS links with GUs, while it is always beneficial under the LoS channel model. As a result, a more complex 3D trajectory optimization problem (as compared to the 2D design in our previous examples under the LoS model) is needed.
Moreover, although the presence of LoS links makes the UAVs well suitable for  5G technologies such as  milimeter wave (mmWave) and massive multiple-input and multiple-output (M-MIMO) communications, the severe air-to-ground interference issue and 3D mobility-induced Doppler effect deserve for more investigations in future work.

Last, for the multi-UAV enabled network, we propose the IUIC as an effective technique to mitigate the strong LoS interference by exploiting the new UAV trajectory design.  Alternatively, motivated by the rapid advance of the wireless backhaul technologies, the UAVs can share messages and perform cooperative beamforming for more efficient  interference mitigation, a technique so-called  coordinate multipoint (CoMP) in the sky \cite{Liu18}. It is worth noting that the
methodology for designing the optimal UAV  trajectories for the CoMP case is generally different from that for the IUIC case. For example, to maximize the cooperative beamforming gain in  CoMP, it may be desirable to let some UAVs form a fleet to serve the GUs along a common trajectory, while this is apparently undesirable in the IUIC case due to the inter-UAV interference. Another important issue worthy of further investigation is how to dynamically adjust the UAV trajectories according to the GUs' movement to improve their throughput and delay performances \cite{Liu18}.

\section{Throughput-Energy Tradeoff}\label{sec:Throughput-Energy Tradeoff}
{In this section, we investigate the joint UAV trajectory and communication deign to characterize the throughput-energy tradeoff.  Specifically, we first discuss the energy consumption models for both  fixed-wing  and rotary-wing UAVs. Then, we revisit the single UAV-enabled system in Section \ref{Sect:singleUAV} by taking into account the UAV's  propulsion energy consumption,  followed by further discussions on related/future work. }
\begin{figure}[!t]
\centering
\subfigure[Fixed-wing UAV.]{\includegraphics[width=0.4\textwidth]{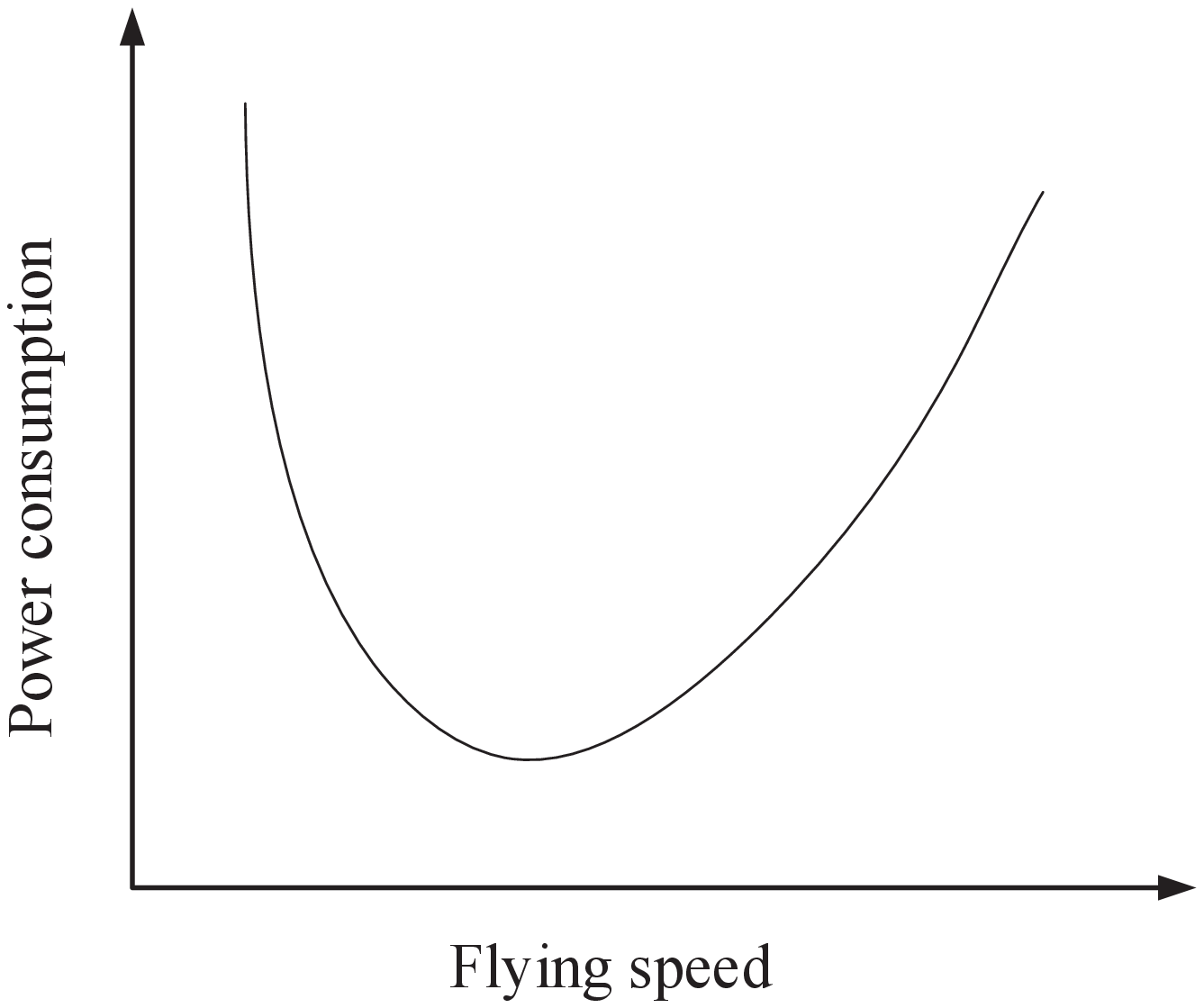}}~~~
\subfigure[Rotary-wing UAV.]{\includegraphics[width=0.4\textwidth]{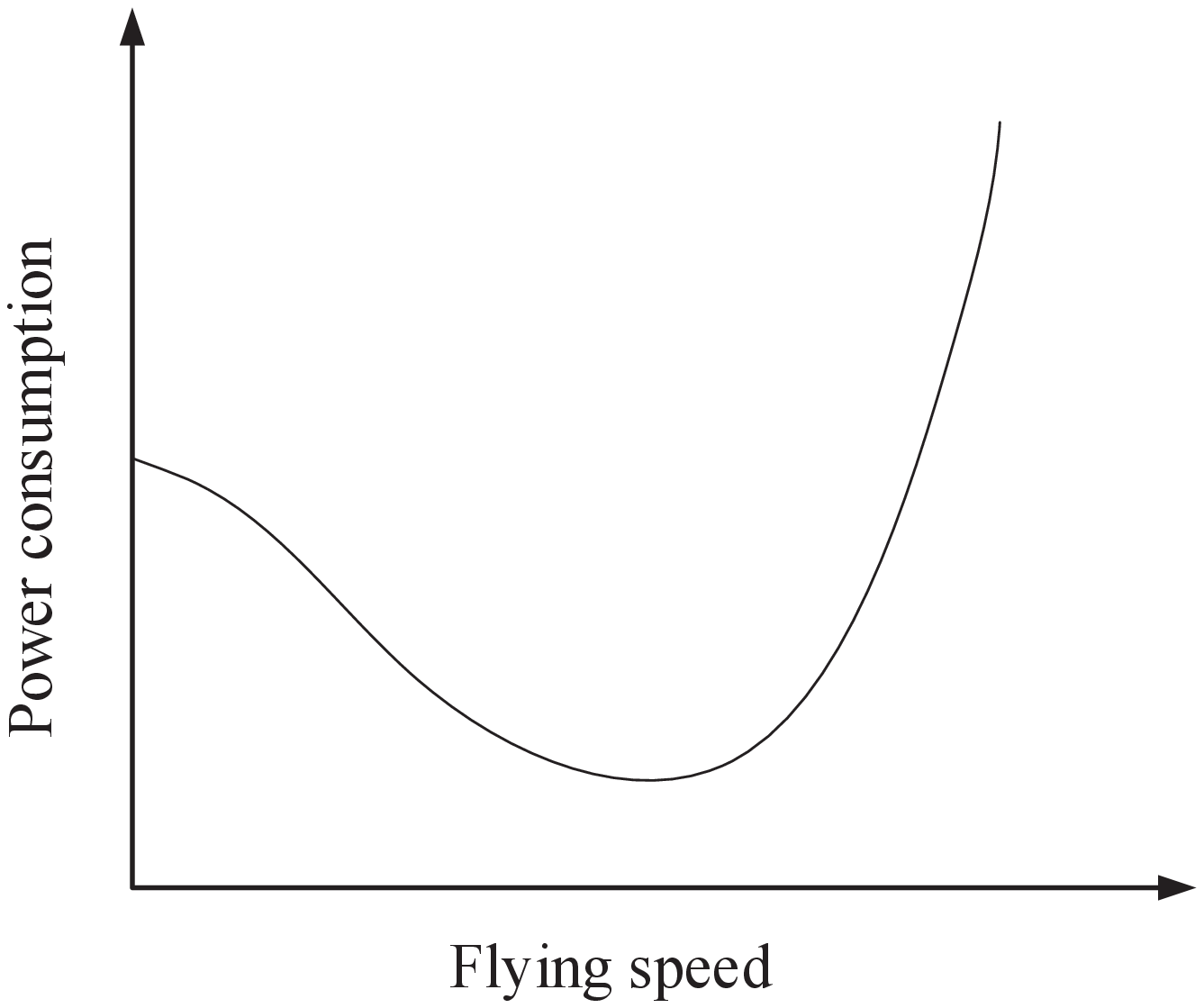}}
\caption{Typical propulsion power consumption versus the UAV's flying  speed.} \label{energy:model} \vspace{-0.5cm}
\end{figure}

\subsection{UAV Propulsion Energy Consumption Model: Fixed-Wing and Rotary-Wing}
{Fixed-wing and rotary-wing UAVs are the two main types of UAVs that have been widely adopted in practice. Both of them possess respective unique sets of advantages and limitations that render them more or less suitable for different applications.
To illustrate the throughput-energy tradeoff, the UAV's propulsion energy consumption needs to be practically quantified  first. Towards this end, two analytical energy consumption models have been established for fixed-wing and rotary-wing UAVs, respectively, in \cite{zeng2017energy} and \cite{zeng2018energy_model}. In general, the propulsion energy consumption of the UAV depends on its velocity (including both the flying speed and direction) as well as the acceleration.  In Fig. \ref{energy:model}, the typical propulsion power consumption versus the UAV's flying speed is illustrated for both fixed-wing and rotary-wing UAVs. In both cases, it can be observed that as the UAV's flying speed increases,  the corresponding  propulsion power consumption first decreases and then increases, which implies that flying at too high and too low speeds (e.g., the spectral-efficient HFH trajectory proposed in Section \ref{Sect:singleUAV}) are neither energy-efficient.
Furthermore, flying at the very low speed is extremely energy-consuming and even impossible for fixed-wing UAVs in practice,  which renders them very difficult to hover over a small geographical area (e.g., as hovering/stationary BSs), while this is not significant for rotary-wing UAVs. However, rotary-wing UAVs suffer from consuming excessive propulsion power when the UAV's flying speed is very high, which makes them not efficient for mission tasks within a wide geographical area (e.g., as high mobile BSs).}

\subsection{Energy-constrained Trajectory Optimization}

\begin{figure}[!t]
\centering
~~~~\,\subfigure[A UAV-enabled two-user wireless system.]{\includegraphics[width=2.7in, height=2in]{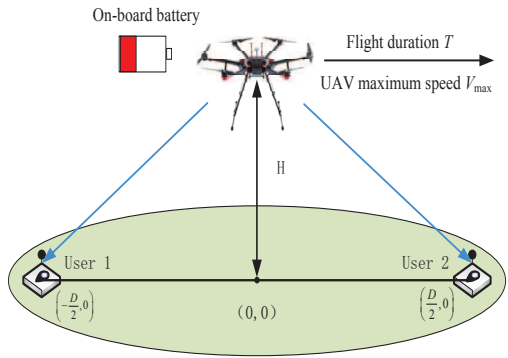}\label{TE:model}}~
\,\,~~\subfigure[UAV's horizontal trajectory for different UAV propulsion energy.]{\includegraphics[width=3.2in, height=2.2in]{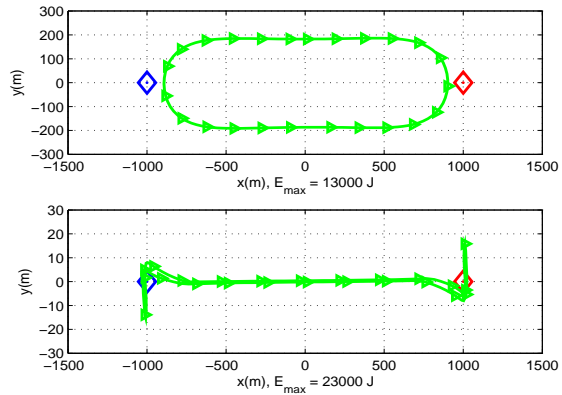}\label{TE:trj}}
\subfigure[UAV's speed over time.]{\includegraphics[width=3.2in, height=2.2in]{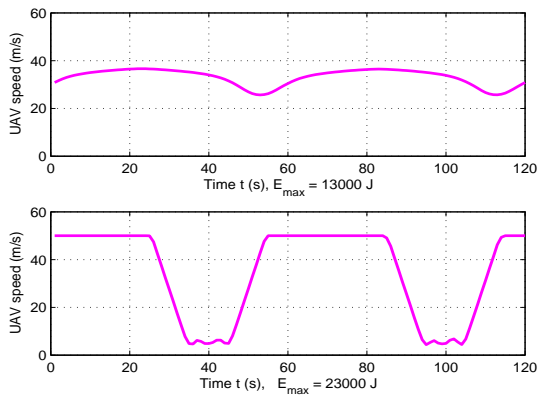}\label{TE:speed}}
\subfigure[Common throughput versus propulsion energy, $E_{\max}$.]{\includegraphics[width=3.2in, height=2.2in]{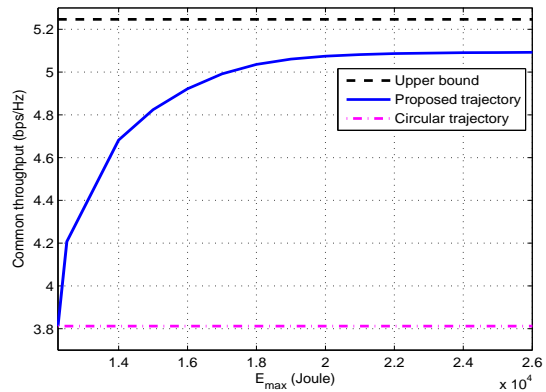}\label{TE:curve}}
\caption{Throughput-energy tradeoff  for a single UAV-enabled network with two GUs.  The GUs' nominal locations are marked by `$\diamondsuit$'s and the UAV trajectories are marked by `$\vartriangleright$'s. For the propulsion power consumption model in \cite{zeng2018energy_model}, the constants $c_1$ and $c_2$ are set as 9.26$\times$10$^{-4}$ and 2250, respectively.
The simulation parameters are set as follows: $V_{\max} = 50$ m/s,  $V_{\min} = 5$ m/s, $a_{\max} = 5$ m/s$^2$, and $T=120$ s. Other parameters are set to be the same as those in Section \ref{Sect:singleUAV}.    } \label{Throughput:energy:tradeoff:singleUAV} \vspace{-0.5cm}
\end{figure}

{As shown in Fig. \ref{TE:model},  we considered the UAV-enabled two-user system where a UAV powered by an on-board battery is dispatched to serve two GUs with a finite flight period $T$. The total amount of available propulsion energy in the battery   is denoted by $E_{\max}$.  For ease of exposition, we consider the fixed-wing UAV in this subsection for which the minimum UAV speed and the maximum acceleration speed are denoted by $V_{\min}$ and $a_{\max}$, respectively.
Other system assumptions such as the UAV's constant flight altitude, quasi-static GUs, LoS channel model, and multiple access scheme are the same as those in Section \ref{Sect:singleUAV}.} {Similarly, we aim to maximize the common throughput between the two GUs via jointly optimizing the UAV trajectory as well as the user scheduling while subject to both the UAV's total energy constraint as well as the mobility constraints.  Due to the high non-convexity of the problem, we apply the same techniques in Section \ref{Sect:singleUAV} to solve the problem and the circular trajectory in adopted for the initialization where the UAV's constant speed is set to achieve the least energy consumption. }

In Fig. \ref{TE:trj}, we plot the UAV's trajectories under different amounts of propulsion energy. It is observed that the UAV tends to fly close to the two GUs by following a smooth trajectory with relatively large turning radii when  $E_{\max}= 13000$ J whereas when this value is increased to  $E_{\max}= 23000$ J, the UAV's trajectory approaches the one that does not take into account the propulsion energy consumption as in Fig. \ref{TD:trj}. This is because sharp changes  on the flying direction as in the latter case although helps reducing the path loss from the UAV to GUs, also requires excessive propulsion energy consumption. Furthermore, the UAV's flying speeds over  time in the above two cases are illustrated in Fig. \ref{TE:speed}. It is observed that in the first case, the UAV's flying speed does not vary much around 30 m/s during the total period due to the limited available propulsion energy while in the latter case with sufficient available propulsion energy, the UAV first flies at the maximum speed (50 m/s) to get close to GUs and then hover around the GU at the minimum seed (5 m/s), so as to fully exploit the controllable mobility to maximize the throughput.  Finally, the achievable throughput versus the propulsion energy is plotted in Fig. \ref{TE:curve}. The upper-bound throughput is obtained by ignoring the propulsion energy constraint, which is exactly the achievable throughput in Fig. \ref{TD:curve} with the same $T$. The lower-bound throughput is the one that achieved by the initial circular trajectory with the UAV's speed being 30 m/s. One can observe that the common throughput can be significantly improved at the cost of propulsion energy consumption. In particular,  as the propulsion energy increases, the common throughput first increases rapidly and then approaches a constant that is strictly lower than the upper-bound throughput.  This is because when the available propulsion energy is sufficiently large (e.g., $E_{\max}=23000$ J) for a given flight period $T$, the throughput will be limited by the practical minimum UAV speed and maximum acceleration speed constraints rather than the propulsion energy.

\subsection{Further Discussion and Future Work}
{The throughput-energy tradeoff can be further extended by taking the GUs' energy consumption into account, for example for UAV-enabled data collection in IoT networks  \cite{yang2017energy,mozaffari2017mobile}.  Since the IoT devices are generally equipped with capacity-limited batteries, how to prolong their life-time is critical for the sustainability and proliferation of the future IoT ecosystem. Thanks to the controllable mobility, the UAV-enabled mobile collector can move sufficiently close to the IoT devices to collect data so as to reduce their transmission energy or the completion time (for collecting a certain amount of data). In the later case,  the energy consumed on circuit operation can be decreased accordingly, which is also critical for energy-limited IoT devices. However, the saved energy at the IoT devices is at the cost of the UAV's movement, which implies an interesting tradeoff between the IoT devices' communication energy consumption and the UAV's propulsion energy consumption \cite{yang2017energy}. 

On the other hand,  the UAVs' energy supply can also  be provided by means of other technologies including solar energy harvesting, laser-beamed wireless power transfer, etc. However, implementing these technologies generally impose additional design considerations that need to be further studied. For example, for solar-powered UAVs, while increasing the flying altitude will lead to higher free space path loss, it helps harvest more solar energy to support more flexible trajectory design (e.g., higher speed and smaller turning radius) for adapting to GUs' real-time movement. As such, a non-trivial tradeoff on optimizing the UAV's altitude exists, which could potentially alleviate the throughput-energy tradeoff in practice.
Furthermore, in the case of mission tasks that employ multiple UAVs, besides the inter-UAV coordination on the signal transmission (e.g., IUIC and CoMP in Section \ref{sec:Throughput-delay Tradeoff}),  novel energy cooperation schemes among the UAVs are also of great practical significance, which includes sequential energy replenishment, wireless power transfer, collaborative trajectory design, etc. For example, the propulsion energy consumption of different UAVs can be balanced via judicious trajectory design so as to maximize the network life-time.



\section{Conclusions}
{In this article, we have revisited the fundamental throughput, delay, and energy tradeoffs in the emerging UAV-enabled wireless communication. In particular, we have shown that the highly controllable UAV mobility with trajectory design can be jointly designed with the communication resource allocation so as to balance the throughput and delay requirements of the users as well as the energy consumption of UAVs. Although this paper focuses on employing UAV as aerial BSs, the proposed tradeoffs are also applicable to the case of UAV users with considerations on throughput, delay, and energy consumption.
It is worth pointing out that besides the three tradeoffs considered in this paper, there exist other design considerations that are worth pursuing. For example, although the communication throughput/delay can be further improved/decreased by employing more UAVs, the cost/deployment efficiency (e.g.,  the number of UAVs and backhaul/fronthaul) may be also of great practical interest. Nevertheless, it is hoped that this article will provide useful insights for the communication and trajectory design of practical UAV-enabled wireless networks and motivate more work in this forward-looking research field.}

\bibliographystyle{IEEEtran}
\bibliography{IEEEabrv,mybib_final}

\end{document}